\documentclass[sigconf]{acmart}
\AtBeginDocument{%
  }

\setcopyright{acmlicensed}

\copyrightyear{2025}
\acmYear{2025}
\setcopyright{acmlicensed}\acmConference[ICS '25]{2025 International
Conference on Supercomputing}{June 8--11, 2025}{Salt Lake City, UT, USA}
\acmBooktitle{2025 International Conference on Supercomputing (ICS '25), June
8--11, 2025, Salt Lake City, UT, USA}
\acmDOI{10.1145/3721145.3735112}
\acmISBN{979-8-4007-1537-2/2025/06}

\def\mycomment#1{}

\usepackage{algorithm}
\usepackage{algpseudocode}
\begin{document}

\title{A3FR: Agile 3D Gaussian Splatting with Incremental Gaze Tracked Foveated Rendering in Virtual Reality}


\author{Shuo Xin}
\additionalaffiliation{%
  \institution{Stanford University, work done during internship at New York University}
  \city{Palo Alto}
  \country{USA}
}
\affiliation{%
  \institution{Tandon School of Engineering \\New York University}
  \city{New York}
  \country{USA}
}

\author{Haiyu Wang}
\affiliation{%
  \institution{Tandon School of Engineering \\New York University}
  \city{New York}
  \country{USA}}

\author{Sai Qian Zhang}
\affiliation{%
  \institution{Tandon School of Engineering \\New York University}
  \city{New York}
  \country{USA}}

\renewcommand{\shortauthors}{Xin et al.}

\begin{abstract}

Virtual reality (VR) significantly transforms immersive digital interfaces, greatly enhancing education, professional practices, and entertainment by increasing user engagement and opening up new possibilities in various industries. Among its numerous applications, image rendering is crucial. 
Nevertheless, rendering methodologies like 3D Gaussian Splatting impose high computational demands, driven predominantly by user expectations for superior visual quality. 
This results in notable processing delays for real-time image rendering, which greatly affects the user experience. Additionally, VR devices such as head-mounted displays (HMDs) are intricately linked to human visual behavior, leveraging knowledge from perception and cognition to improve user experience. These insights have spurred the development of foveated rendering, a technique that dynamically adjusts rendering resolution based on the user's gaze direction. The resultant solution, known as gaze-tracked foveated rendering, significantly reduces the computational burden of the rendering process.

Although gaze-tracked foveated rendering can reduce rendering costs, the computational overhead of the gaze tracking process itself can sometimes outweigh the rendering savings, leading to increased processing latency. To address this issue, we propose an efficient rendering framework called~\textit{A3FR}, designed to minimize the latency of gaze-tracked foveated rendering via the parallelization of gaze tracking and foveated rendering processes. For the rendering algorithm, we utilize 3D Gaussian Splatting, a state-of-the-art neural rendering technique. Evaluation results demonstrate that A3FR can reduce end-to-end rendering latency by up to $2\times$ while maintaining visual quality.
\end{abstract}




\maketitle

\section{Introduction}
\label{sec:intro}
Virtual Reality (VR) technologies fundamentally modify interaction modalities with digital systems, integrating physical and virtual domains through immersive frameworks. Across fields like entertainment, gaming, education~\cite{westin2022inclusive, al2023analyzing,takrouri2022ar}, healthcare~\cite{gerup2020augmented, chengoden2023metaverse, makinen2022user}, and others~\cite{chidsin2021ar, jo2021virtual}, their significance remains substantial. VR facilitates user engagement with sophisticated simulations and scenarios via comprehensive immersive contexts. VR is reshaping not only content consumption but also how we learn, work, and communicate, fostering innovation and expanding possibilities for the future of human-computer interactions. 
 
Image rendering critically determines AR/VR system efficacy and user satisfaction. Within VR, rendering precision substantially affects perceptual authenticity and immersion quality for users. It is vital to achieve high-resolution and low-latency rendering to ensure a seamless and interactive user experience, especially during movements of the head and body. One such rendering algorithm is 3D Gaussian Splatting (3DGS) ~\cite{kerbl20233d} , a rasterization technique for real-time radiance field rendering. It enables the real-time rendering of photorealistic scenes learned from small image samples and achieves state-of-the-art performance in visual quality compared to other methodologies.

\begin{figure}[t]
    \centering
    \includegraphics[width=\linewidth]{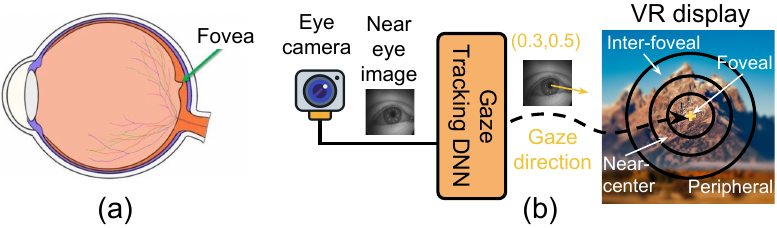}
    \caption{The framework of gaze-tracked foveated rendering.}
    \label{fig:tfr_intro}
\end{figure}


Although the computational power of VR devices has advanced rapidly over the past decade, most rendering algorithms implemented on VR devices, including 3DGS, are still very time-consuming: a major barrier to achieving the desired low-latency, high-resolution output. These complex algorithms, essential for creating detailed and immersive environments, require significant computational resources and processing time, which can disrupt the user experience by causing delays in visual feedback in response to user actions. 

Within VR, the human visual system primarily enables user engagement with virtual environments. Across the visual field, human acuity varies, peaking at the fovea in the retinal center for maximal resolution, as shown in Figure~\ref{fig:tfr_intro} (a). Foveated rendering exploits this acuity gradient, prioritizing computational resources for the central region while reducing allocation to peripheral areas. This technique, termed~\textit{Foveated Rendering}, greatly boosts VR system performance by decreasing the rendering load while maintaining perceived visual quality, establishing it as a vital innovation in VR technology.

To implement foveated rendering, it is essential to track the gaze direction of the users in real time, which then triggers the start of foveated rendering.
Gaze-Tracked Foveated Rendering (TFR) enhances VR rendering efficiency through gaze-tracking systems, typically employing deep neural networks (DNNs) for implementation. By precisely determining the user's point of focus in real-time, TFR system can precisely catch the location of the foveal region which is rendered with the highest resolution (Figure~\ref{fig:tfr_intro} (b)), followed by the \textit{near-center region}, \textit{inter-foveal region} and \textit{peripheral region}, which will be rendered from high to low resolution. 
However, while foveated rendering reduces computational costs, the gaze tracking process adds extra latency that can sometimes exceed the time saved in image rendering.
\begin{figure}[t]
    \centering
    \includegraphics[width=0.95\linewidth]{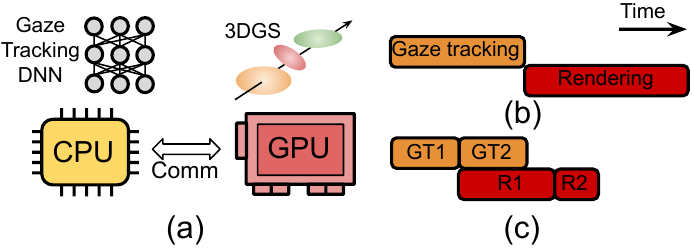}
    \caption{(a) A3FR framework. (b) and (c) depict the workflow of the conventional and A3FR TFR execution, respectively. GT1 and GT2 denote the gaze tracking process, with GT1 representing the completion of generating initial results and GT2 showing the final gaze tracking outcomes. R1 and R2 indicate the incremental rendering of 3DGS.}
    \label{fig:intro1}
\end{figure}

To reduce the overall processing latency of TFR, this paper introduces~\textit{A3FR}, an efficient execution framework for TFR with 3DGS that utilizes parallel processing by executing gaze tracking on the CPU and foveated rendering on the GPU of AR/VR devices, as illustrated in Figure~\ref{fig:intro1} (a). A3FR enables interleaved execution of gaze tracking and rendering tasks, significantly reducing the end-to-end execution time per frame.

To achieve the interleaved operation between gaze tracking and foveated rendering, we have developed a multi-resolution DNN training framework that simultaneously trains the gaze-tracking DNN across different configurations. During the operation, the gaze tracking DNN, implemented on the CPU within the HMD, rapidly produces initial prediction results. These results initially guide the preliminary rendering process of 3DGS on the GPU in the VR system. As the predictions from the gaze tracking DNN become increasingly accurate, the rendering process is incrementally adjusted (Figure~\ref{fig:intro1} (c)), significantly reducing the total execution latency of the conventional TFR, whose processing flow in Figure~\ref{fig:intro1} (b). 

To further optimize the TFR cost of 3DGS, we propose ~\textit{Adaptive Mesh Refinement} (AMR), a technique that reduces computational and memory requirements compared to a uniform resolution while preserving the detailed representation of complex phenomena in advanced simulation and rendering algorithms.
Overall, our contribution can be summarized as follows:

\begin{itemize}
    \item We propose a collaborative execution framework that performs foveated rendering and gaze tracking in parallel, reducing the overall processing latency of AR/VR systems. The framework also adapts to different computational resources by balancing the workload between the CPU and GPU.
    \item We propose an incremental rendering strategy that refines the output over multiple rounds of 3DGS rendering and adaptively adjusts the local resolution based on the visual complexity of the Gaussians. 
    \item To enable incremental gaze prediction, we introduce an efficient gaze-tracking neural network called~\textit{A3FR-ViT}, which supports early gaze direction prediction and facilitates parallel processing for 3DGS rendering.
    \item The evaluation results show that A3FR can reduce the end-to-end rendering latency by over $2\times$ without impacting the user experience, as shown by a comprehensive user study.
\end{itemize}

\section{Background and Related Works}
\label{sec:bg_refs}

\subsection{Oculomotor Behavior in Visual Perception}
\label{sec:bg:human-eye}
\begin{figure}[t]
    \centering
    \includegraphics[width=\linewidth]{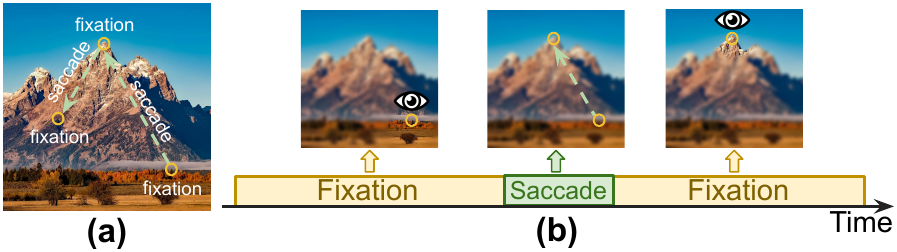}
    \caption{(a) Trace of gaze location. (b) Human eye motion across time.}
    \label{fig:eye_example}
\end{figure}

The human visual system operates through three primary motion modes, each serving distinct roles: \textit{fixation}, which keeps the eye steady on a single point; \textit{saccadic movements}, rapid shifts of gaze between targets; and \textit{smooth pursuit}, which smoothly follows moving objects. Although smooth pursuit is less common, fixation and saccades dominate most visual activities, playing crucial roles in environmental scanning and detail focus, as depicted in Figure~\ref{fig:eye_example} (b). During fixation, the gaze centers on a single point with varying acuity across the visual field. The fovea, located centrally on the retina, provides the highest visual resolution due to its dense concentration of photoreceptor cells, enabling detailed and colorful perception within the direct line of sight. 

Visual acuity diminishes rapidly outside the fovea, resulting in less sensitivity to fine details in peripheral vision. Humans typically perform one to three rapid saccadic eye movements per second~\cite{kowler2011eye, fabius2019time, kwak2024saccade}, each lasting about 20--100 ms and reaching speeds over 200$^\circ$ per second~\cite{robinson1964mechanics}. These swift movements cause temporary visual blur, known as saccadic blur~\cite{matin1974saccadic, campbell1978saccadic}. Figure~\ref{fig:eye_example} (a) illustrates gaze behavior within a scene, showing fixation points connected by saccades before transitioning to another scene. This study focuses on optimizing rendering costs during fixation, which constitutes the majority of viewing time.

\begin{figure*}[t]
    \centering
    \includegraphics[width=\linewidth]{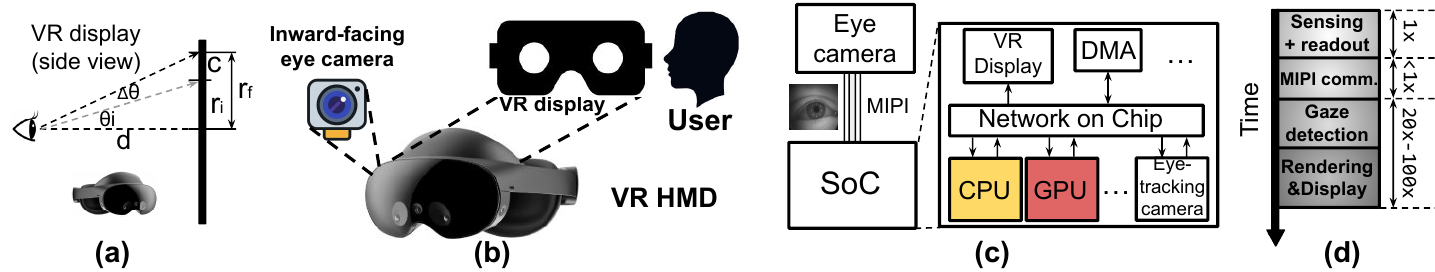}
    \caption{(a) Relationship between tracking error and foveal region size. (b) Meta Quest Pro Head-Mounted Display (HMD). (c) Hardware system layout of VR device. (d) Latency breakdown of TFR for a single frame.}
    \label{fig:combined}
\end{figure*}
\subsection{Foveated Rendering}
\label{sec:bg:tfr}
Based on the user's visual behavior, rendering resources can be dynamically allocated. The TFR method employs DNN to process images from the HMD's internal eye-facing cameras, precisely identifying where the user is focusing at any given moment. The system then creates a perceptually-optimized rendering hierarchy with the \textit{foveal region} rendered at maximum resolution, surrounded by the \textit{near-center region}, \textit{inter-foveal region}, and \textit{peripheral region} with progressively reduced detail~\cite{foveated_rendering_basic, latency_effect_3, ye2022rectanglefr}, as illustrated in Figure~\ref{fig:tfr_intro} (b). Figure~\ref{fig:combined} (a) demonstrates how the high-resolution foveal regions are determined. The radius $r_f$ of this zone is:
\begin{equation}
\label{eqn:foveal_region}
r_f = r_i + c = \rho d \cdot \tan(\theta_i + \Delta\theta) = \rho d \cdot \tan(\theta_f)\footnote{This formula assumes the gaze is centered at the front view, representing the maximum radius of the rendering region.},
\end{equation}
where $\rho$ represents the display's pixel density, $d$ is the fixed viewing distance in HMD configurations, $\theta_i$ corresponds to the eccentricity angle of high-acuity vision, and $\Delta\theta$ accounts for gaze tracking measurement error. $r_i = \rho d \cdot \tan(\theta_i)$ defines the ideal high-resolution area, while $c = \rho d \cdot [\tan(\theta_i + \Delta\theta) - \tan(\theta_i)]$ provides a safety margin to maintain visual quality despite tracking uncertainties. Research by Lin et al.~\cite{lin2024metasapiensrealtimeneuralrendering} established an optimal $\theta_i$ value of 18° for contemporary VR applications. In most HMD settings, the viewing distance $d$ remains constant due to the fixed position of the display relative to the user's eyes. 

According to equation~\ref{eqn:foveal_region}. a significant gaze tracking errors would require larger high-resolution rendering zones, potentially undermining the computational benefits of foveated rendering. Processing additional high-resolution pixels substantially increases rendering demands and power consumption. Therefore, precise gaze tracking becomes essential for maximizing system efficiency. Prior studies by Lavalle et al.~\cite{lavalle2014head} and Kress et al.~\cite{kress2020optical} demonstrate that maintaining a seamless visual experience requires total system latency between \textbf{20-50ms}~\cite{albert2017latency}, emphasizing the importance of optimizing both tracking accuracy and rendering performance.

\subsection{Neural Rendering}
\label{sec:bg:neural_rendering}
3D reconstruction and novel view generation are essential tasks in computer vision and graphics. Recent years have seen the emergence of neural rendering techniques, which generate high-quality images by inference from a learned neural network. Among these, Neural Radiance Fields (NeRF)~\cite{NeRF} has gained significant attention for its ability to generate photorealistic images from 3D scenes. However, NeRF is computationally expensive. Several variants \cite{NeRF-DS, Mip-NeRF360, Mip-NeRF, Zip-NeRF} have been proposed to reduce rendering latencies while preserving visual quality, but they still face challenges in achieving real-time performance.

\begin{figure}
    \centering
    \includegraphics[width=\linewidth]{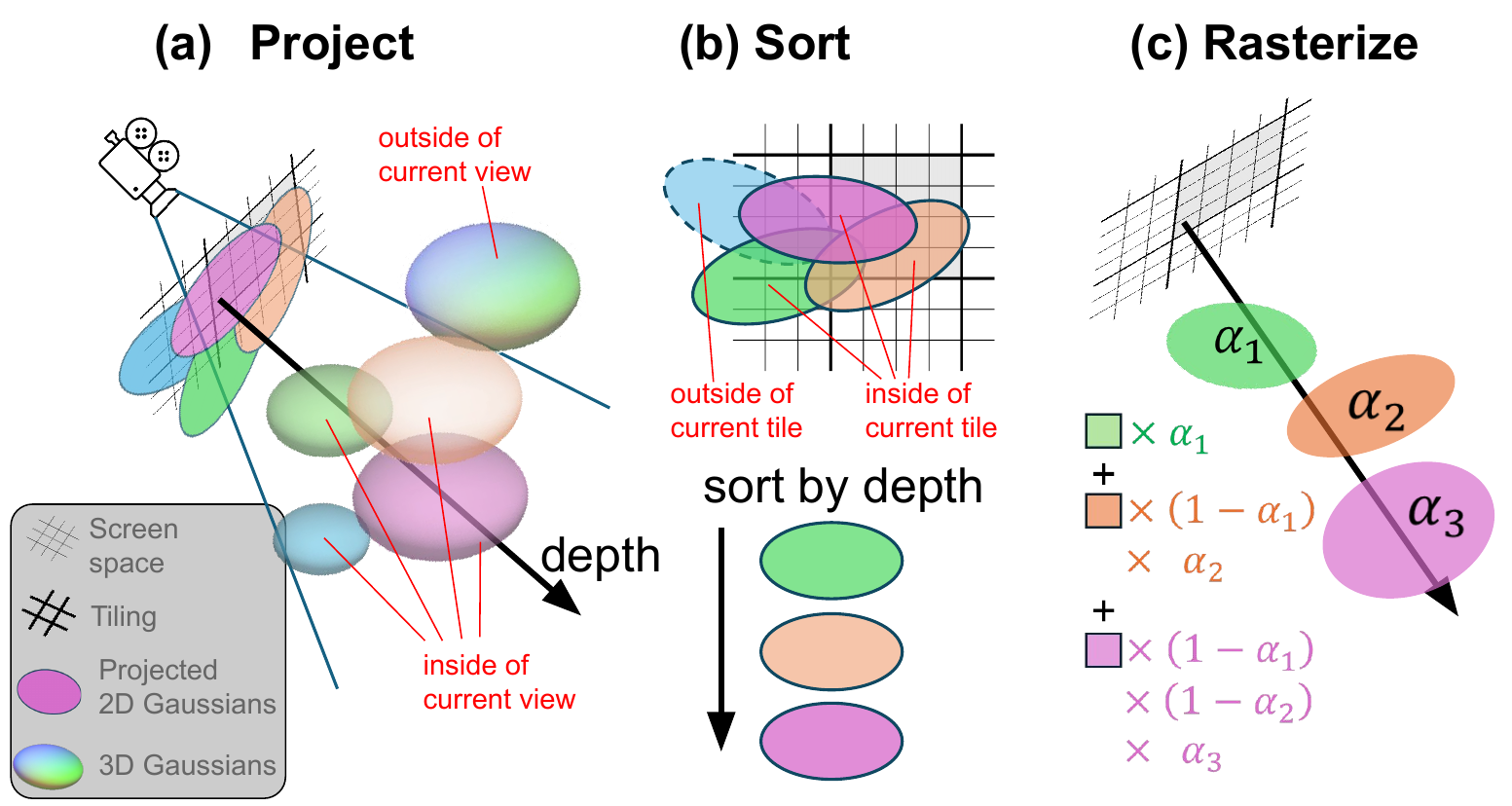}
    \caption{An example of 3DGS Process. (a) Given a camera pose, all 3G Gaussians (colored ellipsoids in figure) that are inside the current camera view are projected to screen space. (b) The projected 2D Gaussians on screen space (colored ellipses in figure) that are inside the current tile are sorted by depth, i.e. distance to camera screen. (c) The opacities $\alpha_i$ of the relevant Gaussians are computed. The Gaussians deposit their color one by one in the previously sorted order. }
    \label{fig:3DGS_demo}
\end{figure}

Recently, an alternative approach, 3D Gaussian Splatting~\cite{kerbl20233d} (3DGS), has been shown to significantly reduces the computational cost of the rendering process. Gaussian Splatting represents scenes using a point cloud, with each point endowed with trainbale position, color, opacity, angular distribution, and radial Gaussian distribution for rasterization.
An example is shown in Figure~\ref{fig:3DGS_demo}. 
The rendering begins with a model trained offline, representing the scene as a point cloud. Each point corresponds to an ellipsoid, shaped by the scales of 3D Gaussian distributions—referred to as a Gaussian point. These ellipsoids are equipped with trainable parameters that control their scales, positions, orientations, opacity, and color distribution, the latter of which is parameterized using Spherical Harmonics (SH). Once the trained points and ellipsoids are established, the online rendering process consists of three primary steps:~\textit{Projection},~\textit{Sorting}, and ~\textit{Rasterization}. Each step plays a critical role in transforming the 3D model data for final image production on screen.

During the projection stage (Figure~\ref{fig:3DGS_demo} (a)), each ellipsoid is initially projected as an ellipse onto the image plane. The process involves determining which ellipses intersect with a given pixel tile (for instance, {a 4$\times$4 area as shown by the shaded region in the figure}) to accurately contribute to the pixel colors within that tile. After that, for each tile, intersecting ellipses are sorted by their depth relative to the image plane (Figure~\ref{fig:3DGS_demo} (b)). This sorting prioritizes ellipses closer to the camera screen, allowing them to have greater influence on the calculation of pixel colors.

Finally, the intersections of all ellipses within a tile and each pixel are calculated. The color of a pixel is then determined using the classical volume rendering method. This method sums the contributions of all intersecting ellipses, processing them from the nearest to the farthest. This technique ensures that the pixel color accurately reflects the visual depth of the scene, as described in Figure~\ref{fig:3DGS_demo} (c). Note that, since the images are rendered at the granularity of tiles, it is possible to selectively process only a portion of pixels on each tile. This allows us to make a tradeoff between rendering resolution and GPU workload.


\subsection{Eye-tracking}
\label{sec:bg:gaze_tracking}
Eye-tracking methodologies can be classified into two principal categories: appearance-based and model-based techniques~\cite{Overview_tracking, Over_tracking_1}.

Appearance-based tracking techniques establish direct relationships between eye imagery and gaze directions~\cite{Overview_tracking, appearance_zhang_1, appearance_zhang_2, liu2025fovealnet, wanghardware}. These methods necessitate extensive training datasets and have catalyzed innovations across various machine learning frameworks, including linear regression models~\cite{linear_regression_2}, random forest algorithms~\cite{random_forest_1}, k-nearest neighbor approaches~\cite{knn_1}, and convolutional neural networks~\cite{resnet_inception,mazzeo2021openeds}. Vision Transformers~\cite{dosovitskiy2020image}, which have achieved remarkable results across numerous computer vision applications, have also been integrated into gaze tracking systems~\cite{nvgaze}. However, their substantial computational demands present considerable obstacles for deployment in real-time tracking environments.

In model-based approaches, researchers employ three-dimensional eye models that replicate the eye's biological structure to determine gaze vectors~\cite{Model_based_method, Model_based_method_1, Model_based_method_2, lu2022model}. The implementation typically involves a dual-stage process: initially extracting critical eye features through specialized neural networks and mapping these onto geometric models, followed by calculating the gaze direction from this representation. This methodology effectively transforms gaze tracking into a segmentation task, frequently utilizing convolutional U-Net architectures, which account for the majority of computational resources~\cite{vrpaper_zhu, deepvog, etracker, zhang2024swifteye}. While previous research has demonstrated excellent accuracy in eye segmentation~\cite{RITnet}, the subsequent gaze direction estimations can deviate by more than $2^\circ$ from actual measurements. These discrepancies primarily stem from imprecise eye center and radius calculations during initialization, alongside limitations inherent to geometric models during optimization procedures, creating misalignment between estimated and actual gaze patterns~\cite{Model_based_method}.


\section{Methods}
\label{sec_methods}
\subsection{Overview}
\label{sec:preliminary}
A typical VR device, such as a HMD, is shown in Figure~\ref{fig:combined} (b). The inward-facing eye camera of the VR device continuously captures images of the eye to facilitate the execution of TFR. The foveated rendered scene is then displayed on the screen for the user to see. Figure~\ref{fig:combined} (c) shows the high-level architecture of TFR system, which consists of three primary components: a near-eye camera (image sensor), a host processor, and a interconnection link (MIPI~\cite{lancheres2019mipi}). The standard TFR workflow is depicted in Figure~\ref{fig:combined} (d). Initially, an eye image is captured by a monochrome camera situated near the eye. This image undergoes preprocessing and readout by the image signal processor (ISP) before it is transmitted to the host processor via the MIPI link. Upon reception, the host processor forwards the image to an eye-tracking DNN, which determines the gaze direction. The gaze direction then informs the foveated rendering process, which adjusts the rendering of the VR scene accordingly. 

Figure~\ref{fig:combined} (d) provides an approximate latency breakdown of the conventional TFR process. Camera sensing latency $T_{s}$ and MIPI communication delay $T_{c}$ approximately 1 ms~\cite{you2023eyecod,liu20204,angelopoulos2020event,sun2024estimating, zhao2024neural} and less than 1 ms~\cite{lee20196,mipi}, respectively, accounting for a small fraction of the total latency. In contrast, the gaze detection latency $T_{d}$, along with the subsequent rendering and display process $T_{r}$, usually consumes a much larger portion (20-100$\times$ longer) of the overall latency, based on the studies from~\cite{albert2017latency, power_quality_tradeoff}. It is evident from Figure~\ref{fig:combined} (d) that the majority of the total processing time $T_{tot} = T_{s} + T_{c} + T_{d} + T_{r}$ is consumed by gaze detection $T_{d}$ and rendering $T_{r}$. Most TFR frameworks in HMDs execute the gaze tracking and foveated rendering process sequentially on the GPU within the VR device~\cite{quest_pro_tfr, singh2023power, patney2016perceptually}, which results in the underutilization of other computational resources (e.g., CPU shown in Figure~\ref{fig:combined} (c)). 

To illustrate this, Table~\ref{tab:latency_examples} presents the latency of two gaze tracking DNNs, as introduced in~\cite{fovealnet2025liu}, with three and six layers, respectively. These DNNs are tested on the CPU and GPU of the Nvidia Jetson Orin NX, a commonly used platform to simulate VR devices~\cite{gilles2023holographic, zhang2024boxr, zhang2024co, pancrisp}. Additionally, we simulate the 3DGS rendering process at three resolutions: 720P, 1080P, and 1440P on the GPU of the Jetson Orin NX. We note that running eye tracking concurrently with the 3DGS introduces an additional 12ms of TFR latency. In contrast, offloading the gaze tracking task to the CPU results in a latency similar to that of 3DGS rendering on the GPU.


\begin{table}[t]
    \centering
    \label{}
    \begin{tabular}{cccc}
    \toprule
         &GPU (6 layers) & {CPU (3 layers)} & {CPU (6 layers)}  \\
         ViT latency& 11.43 ms& 16.44 ms & 30.93 ms \\
    \toprule
    \toprule
    & GPU(720P) & GPU(1080P) & GPU(1440P) \\
    3DGS latency & 27.00 ms & 37.78 ms & 53.14 ms\\
    \toprule
    \end{tabular}
    \caption{Latency of ViT eye-tracking and 3DGS rendering}
    \label{tab:latency_examples}
\end{table}

Therefore, it would be advantageous to parallelize the gaze tracking process with the 3DGS rendering process, utilizing both the CPU and GPU to minimize the total TFR latency, denoted as \( T_{tot} \). Thus, \( T_{tot} \) becomes \( T_{tot} = T_{s} + T_{c} + \max(T_{d}, T_{r}) \).

The results in Table~\ref{tab:latency_examples} indicate that it would be advantageous to parallelize the gaze tracking process with the 3DGS rendering process, utilizing both the CPU and GPU to minimize the total TFR latency, denoted as \( T_{tot} \). Thus, \( T_{tot} \) becomes \( T_{tot} = T_{s} + T_{c} + \max(T_{d}, T_{r}) \). To accomplish this, A3FR offloads the less demanding gaze tracking DNN to the CPU, which allows the GPU to dedicate its resources to the more demanding rendering tasks. Furthermore, we introduce A3FR-ViT (Section~\ref{sec:multi-resolution-training}) to facilitate an early exit capability for the gaze tracking DNN. This feature produces preliminary gaze tracking results that are immediately sent to the GPU to start the foveated rendering process. As the gaze tracking continues, increasingly accurate results are generated, allowing for ongoing enhancements in the rendering quality. This approach significantly reduces overall latency, as depicted in Figure~\ref{fig:intro1} (c). Next, we describe the design of A3FR-ViT in Section~\ref{sec:multi-resolution-training} and A3FR intremental rendering scheme in Section~\ref{sec:a3fr_scheme}.

\subsection{Design of Gaze Tracking Neural Network}
\label{sec:multi-resolution-training}

\begin{figure}[t]
\centering
\includegraphics[width=\linewidth]{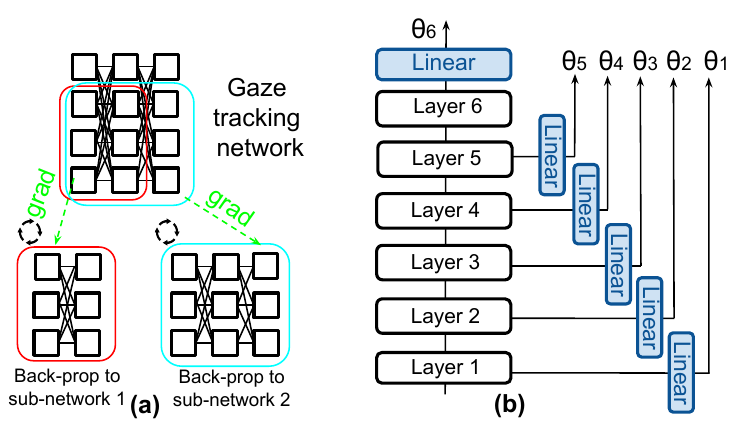}
\caption{(a) Multi-resolution training framework. (b) An example of N=6 layer ViT with early-exit mechanism, the output from each layer is connected to a linear layer to produce the gaze prediction. 
}
\label{fig:multi-resolution-training}
\end{figure}

Vision transformer (ViT)~\cite{dosovitskiy2020image} has been shown to perform well on various tasks in computer vision. In this work, we use multilayer ViT~\cite{dosovitskiy2020image} for eye-tracking, where the input is the image of the user's eye captured by VR camera and the output is the user's gaze direction. The resultant ViT, termed~\textit{A3FR-ViT}, processes the input image by segmenting it into patches, tokenizing the patches, and adding positional data before passing them through the transformer block. The architecture includes 6 transformer blocks, each featuring 6 heads and an embedding dimension of 384. Modifications to the original ViT design replace the classifier MLP layers with a series of linear layers, which output the 2D gaze direction.

To achieve parallel operation between gaze tracking and foveated rendering, we train the ViT so that the parameters in intermediate layers are used to give early predictions, which are used for foveated rendering, before all layers finish. Specifically, we design a multi-resolution DNN training strategy that simultaneously optimizes several sub-networks across distinct DNN architectures (Figure~\ref{fig:multi-resolution-training} (a)). This joint optimization framework yields a multi-resolution model that operates at varying depths. Initially, preliminary gaze tracking results are generated, which are then utilized by the foveated rendering process. Subsequently, the two processes run in parallel, enhancing the efficiency and responsiveness of the system.

To implement this, we add a linear layer to the end of the selected encoder blocks within the A3FR-ViT, allowing it to generate a gaze direction prediction based on the intermediate outputs, as depicted in Figure~\ref{fig:multi-resolution-training} (b). Here, let $N$ represent the total number of selected layer blocks in the A3FR-ViT that are appended with local exit. This configuration yields a series of $N$ predictions on the gaze location, $u_{n}$, derived from the intermediate results, where $1\leq n \leq N$. $u_{n} = \{u_{n,x}, u_{n,y}\}$ further consists of x and y coordinate of the VR display. Consequently, the loss function $L_{f}$ for the multi-resolution training is defined as the sum of the training losses from each of these early-exit points.
\begin{equation}
    \label{eqn:multiresolution-loss}
    L_{f} = \sum_{n=1}^{N} \lambda_{n} L_{g} (u_n, u_{gt}),
\end{equation}
where \( u_{gt} \) denotes the ground truth gaze location in the dataset, and \( \lambda_{n} \) is a hyperparameter that specifies the importance of gaze predictions at each early-exit point. \( L_{g} \) represents the gaze prediction loss function, with \( L_2 \) loss being utilized in this study. The parameter \( \lambda_{n} \) reflects the relative weight assigned to each loss function. $u_{N}$ denotes the final gaze prediction.

To further minimize the computational demands of A3FR-ViT, we implement tokenwise pruning~\cite{spvit} on the input tokens by evaluating their significance (attention scores) in relation to the final gaze prediction and discarding the less important tokens. In the self-attention mechanism of the model, tokens undergo a linear transformation into Query, Key, and Value matrices. The attention score is calculated through a dot product between the Query and Key matrices, which is then scaled and processed through a Softmax operation. This score reflects each token's relevance to the gaze prediction outcome. Using these scores, tokens with an attention score below a specified threshold, \(\sigma\), are filtered out. This pruning effectively reduces the computational load on subsequent ViT blocks by lowering the number of input tokens and, consequently, shrinking the size of the intermediate activations.
\begin{figure*}
    \centering
    \includegraphics[width=1\linewidth]{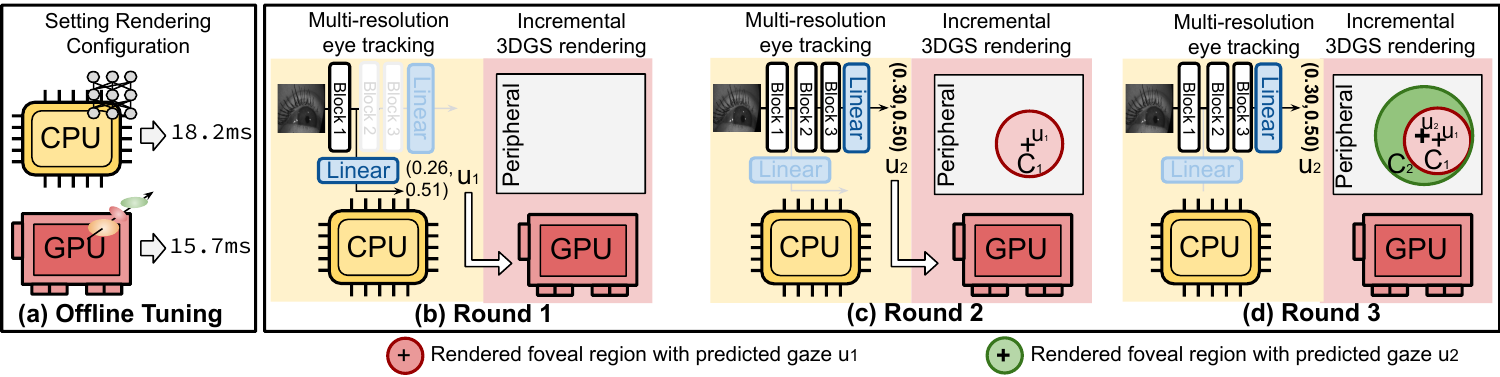}
    \caption{Illustration of the proposed parallel TFR procedure for a simple eye-tracking model only 1 early exit. In round 1, the 3DGS model starts precomputing and rendering at peripheral resolution. In round 2, 3DGS model receives an intermediate gaze prediction and refines pixels the foveal regions. In round 3, final gaze prediction is passed and 3DGS model makes corrections using the more accurate gaze. }
    \label{fig:render_rounds}
\end{figure*}

\subsection{A3FR Incremental Rendering Scheme}
\label{sec:a3fr_scheme}

As discussed in Section \ref{sec:multi-resolution-training}, the ViT eye-tracking system provides a series of gaze predictions \( u_{i} \) at each exit point. These predictions become more accurate progressively. The A3FR system begins foveated rendering using these early predictions even before the final gaze location is confirmed, refining the rendering focus as more precise predictions become available. This strategy leverages early predictions effectively, reducing unnecessary processing and wait times. In terms of system operations, the A3FR-ViT and 3DGS rendering processes run simultaneously on the CPU and GPU within the HMD, enhancing hardware utilization and reducing total computational delay for TFR. Prior to execution, the CPU and GPU latencies on gaze tracking neural network execution and foveated rendering are first profiled offline (Figure~\ref{fig:render_rounds} (a)), and a subset of \( M \) exit points (\( M < N \)) is selected from the \( N \) available local exit points to ensure optimal synchronization between the CPU and GPU. Figure~\ref{fig:render_rounds} illustrates an example where $M=2$.
While A3FR-ViT works on its initial layers, 3DGS concurrently renders the peripheral areas which do not require precise gaze information as illustrated in Figure~\ref{fig:render_rounds} (b). When the first gaze prediction \( u_{1} \) is made, it is sent to the GPU to adjust the rendering to higher resolutions as needed, shown in Figure~\ref{fig:render_rounds} (c). With each subsequent, more accurate prediction \( u_{1}, u_{2} \), etc., 3DGS adjusts its rendering area to align with the new gaze data, as depicted in Figure~\ref{fig:render_rounds} (d).


\subsubsection{Incremental Rendering Strategy}
\label{sec:rendering_setting}
\begin{figure}
    \centering
    \includegraphics[width=1\linewidth]{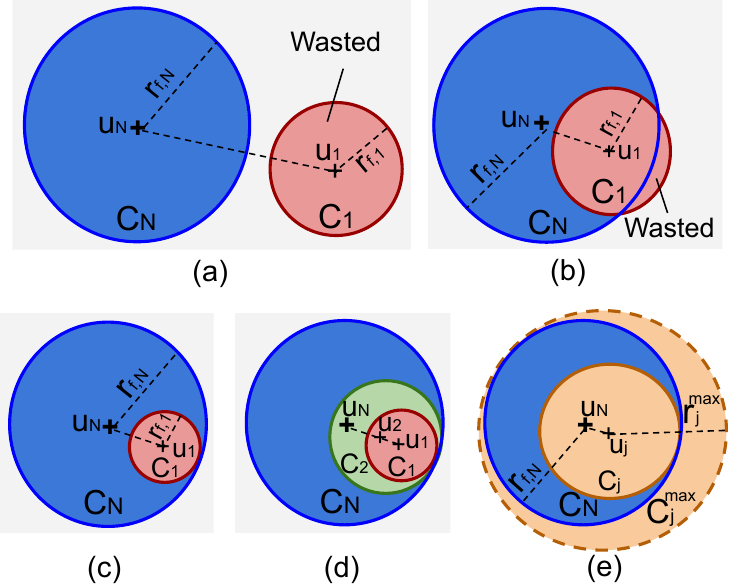}
    \caption{An illustration of the A3FR operation is provided: (a) If the distance between \(u_{1}\) and \(u_{N}\) is large, the entire region centered at \(C_{1}\) will be wasted. (b) As \(u_{1}\) and \(u_{N}\) get closer, the extent of the wasted region decreases. (c) No region will be wasted if the radius \(r_{f,1}\) meets the requirements set forth in equation~\ref{eqn:criteria1}. (d) The green region will be rendered as \(u_{2}\) approaches \(u_{N}\). (e) If no additional gaze predictions are forthcoming, the rendering process will persist until the rendering of \(C^{max}_{j}\) is complete, indicated by the dotted yellow circle. }
    \label{fig:example}
\end{figure}

Due to potential errors in initial gaze predictions, the areas rendered in high resolution based on early results (e.g., \( u_{1} \)) might not correspond to the actual regions requiring high-resolution detail. For simplicity, we consider the scenario where the scene is rendered at two levels of resolution—one for the foveal area and another for the peripheral area. However, the proposed incremental rendering strategy can be adapted to accommodate multiple resolution levels. An example is illustrated in Figure~\ref{fig:example} (a): Suppose the preliminary gaze prediction \( u_{1} \) is initially sent to the GPU, triggering the rendering process for the region \( C_{1} \) with a radius of \( r_{f,1} \) at the highest resolution. If \( u_{1} \) is significantly far from the final predicted gaze location \( u_{N} \), this discrepancy results in substantial redundant rendering since \( C_{1} \) does not align with the actual foveal region \( C_{N} \), determined by \( u_{N} \) with a radius $r_{f,N}$. Conversely, when \( u_{1} \) is sufficiently close to \( u_{N} \), the overlap between \( C_{1} \) and \( C_{N} \) increases, and the wasted rendering region gets smaller, as depicted in Figure~\ref{fig:example} (b). 

Our aim is to \textbf{minimize unnecessary rendering computations} by ensuring that the areas rendered based on initial gaze predictions (e.g., \( C_{1} \)) are effectively used in subsequent refinement processes. This approach optimizes the use of computational resources by leveraging the initial rendering output in later, more detailed rendering stages. Specifically, to eliminate the unnecessary rendering in \( C_{1} \), we must ensure that the sum of the radius of \( C_{1} \) and the distance between \( u_{1} \) and the final gaze prediction \( u_{N} \) is less than the foveal region radius \( r_{f,N} \) denoted by \( C_{N} \) (Figure~\ref{fig:example}(c)). This condition guarantees that \( C_{1} \) is entirely contained within the foveal region \( C_{N} \). Mathematically, this relationship is formalized in the following theorem:
\begin{equation}
\label{eqn:criteria1}
r_{f,i} \leq max(0, r_{f,N}-dist(u_{i}, u_{N}))  \enspace\enspace 1\leq i\leq N
\end{equation}
where i is set to 1 shown in Figure~\ref{fig:example} (c) and $dist(.)$ denotes the $l_2$ distance between two points. To ensure efficient utilization of computational resources, the maximum radius of the rendered region should be determined by Equation~\ref{eqn:criteria1}. Once the next predicted gaze location \( u_{i} (1\leq i \leq N) \), is generated and is closer to \( u_{N} \), the corresponding radius \( r_{f,i} \) can be repeatedly computed using Equation~\ref{eqn:criteria1}, ensuring that only the incremental region is rendered, as shown in green in Figure~\ref{fig:example} (d).

\subsubsection{Offline Profiling}
\label{sec:offline_profile}
To optimize rendering settings, equation~\ref{eqn:criteria1} is based on the known distances \( \text{dist}(u_{i}, u_{N}) \) between the predicted gaze locations \( u_{i} \) and \( u_{N} \). In practical applications, we utilize the average distance between gaze locations as determined from eye-tracking training data (e.g. OpenEDS 2020~\cite{OpenEDS2020}). Consequently, Equation~\ref{eqn:criteria1} can be reformulated to incorporate this average distance, providing a more empirical basis for adjusting rendering parameters to enhance accuracy and efficiency in real-time applications.
\begin{equation}
\label{eqn:criteria1}
r_{f,i} = max(0, r_{f,N}-\mathbb{E}(dist(u_{i}, u_{N})))  \enspace\enspace 1\leq i\leq N
\end{equation}
where \( \mathbb{E}(.) \) denotes the expected value across the training data set. 

\subsubsection{Incremental Rendering under Dynamic Performance Variation}
\label{sec:dynamic_rendering}
To minimize overall latency, it is crucial that the CPU completes the gaze prediction at the \(i\)-th local exit $L_{i}$ before the rendering of \(C_{i-1}\) finishes. If not, \(C_{i}\) will have to wait until \(u_{i}\) is fully processed, resulting in increased overall latency. Additionally, the processing speeds of the GPU and CPU can fluctuate over time due to resource sharing with other programs, making it challenging to develop a fixed scheduling algorithm for minimal processing time.

To address this issue, we've developed an asynchronous incremental rendering scheme that allows the GPU and CPU to operate independently. Under this scheme, the latest predicted gaze \( u_{i} \) is stored in GPU memory, and the incremental rendering process runs concurrently on the GPU. Once the GPU completes rendering the current region \( C_{j}, (j < i) \), it retrieves the most recent gaze prediction from its memory and initiates the next round of rendering based on this updated gaze information, as depicted in Figure~\ref{fig:render_rounds}.

However, one issue that arises is when the CPU processing speed significantly lags behind that of the GPU, potentially delaying the availability of the next gaze prediction \( u_{i} \) when \( C_{i-1} \) finishes rendering. This situation can prevent the GPU from proceeding with the next round of the rendering process. To address this, we have developed a~\textbf{speculative incremental rendering} (SIR) method, which allows the rendering process to continue even if no new gaze prediction is available. 
In this scenario, the incremental rendering process will continue, centered on \(u_{j}\), until a more accurate gaze prediction is received. If no additional gaze predictions are received, the rendering process will conclude once the region \(C^{max}_{j}\), defined by the maximum radius \(r^{max}_{j}\), has been fully rendered. This can be illustrated as follows:
\begin{equation}
\label{eqn:max_radius}
    r^{max}_{j} = r_{f,N}+\mathbb{E}[dist(u_{j}, u_{N})] 
\end{equation}

An example is shown in Figure~\ref{fig:example} (e). Assume that \(C_{j}\) has been rendered and no further gaze predictions are received. In this case, the rendering process will continue until the region delineated by a dotted circle, with a radius of \(r^{max}_{j}\), is fully rendered. This ensures coverage of the actual foveal region centered at \(u_{N}\).

\subsection{AMR-based Rendering Strategy of A3FR}
\label{sec:AMR}
Adaptive mesh refinement (AMR) is a technique widely used in scientific computing and finite-element methods \cite{Berger:1984zza,loffler2012einstein,zhang2021amrex}. It adaptively reduces/enhances the local tile resolution based on local error estimation in a scientific simulation or graphic rendering. AMR saves the computation workload and memory usage, compared to a uniform resolution, while maintaining the details of various phenomena in intricate simulation and rendering algorithms. Moreover, it allows one to reuse the results from lower quality levels when computing higher levels of refinement. In this paper, we further apply this idea to the practice of 3DGS.

\begin{figure}
    \centering
    \includegraphics[width=1\linewidth]{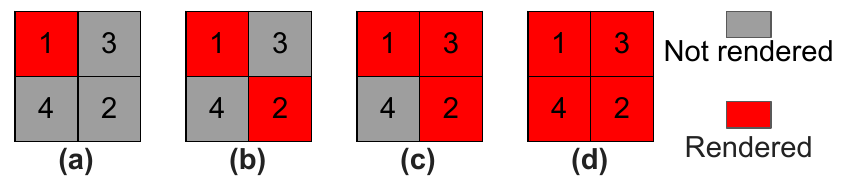}
    \caption{The adaptive rendering strategy of A3FR involves varying the rendering detail based on proximity to the gaze location. For pixel tiles in the peripheral region, only the top left corner is rendered. As the tiles get closer to the gaze location, more pixels within each tile are rendered.}
    \label{fig:AMR-demo}
\end{figure}

Building on the incremental rendering scheme outlined in Section~\ref{sec:rendering_setting}, we now explore adaptive adjustments to the rendering resolution for 3DGS to achieve computational savings. Specifically, for {each block of adjacent $2\times 2$ pixels}, depending on the relative distance between the pixel tile and the predicted gaze direction, a subset of the pixels within the current tile will be rendered accordingly. As illustrated in Figure~\ref{fig:AMR-demo}, for each block of adjacent $2\times 2$ pixels in the peripheral region, only the top left pixel is rendered. Similarly, in the inter-foveal, near-center, and foveal regions, the number of rendered pixels within the $2\times 2$ tile increases accordingly. The un-rendered pixels simply inherit or interpolate from the rendered pixels. This provides a knob of 4 levels of resolutions, ideally corresponding to 4 regions of eccentricity from gaze center. Our implementation can also be extended to any $N^2$ levels, by selectively rendering pixels on $N\times N$ block. The user study described in Section~\ref{sec:user_study} shows that this adaptive rendering strategy will not cause appreciable visual quality degradation, while greatly saves the computational cost.

{
\subsection{Implementation Details}
In practice, to implement incremental rendering, we made several technical changes to the original 3DGS algorithm~\cite{3D-GS}. First, before rendering, the 3DGS algorithm precomputes several intermediate quantities: 3D Gaussian points projected to screen space, subsets of Gaussian points on each tile, sorted list of Gaussians by depth and etc. To achieve rendering by multiple rounds without redundant precomputing, the intermediate variables computed in round 1 are buffered, and later rounds directly use the buffered variables without the need to access the original 3D Gaussian representations of the scene. Also, to make the best use of the \textit{cooperative group} feature in CUDA, the original 3DGS model uses a tiling of $16\times 16$ pixels per tile, i.e. 256 threads per cooperative group at render time. To achieve adaptive, incremental rendering, we make the following changes: we choose a tiling of $32\times 32$ pixels per tile. The tile is further subsampled to $16\times 16$ blocks of $2\times 2$ pixels. At render time, the tile will raise 4 cooperative groups, corresponding to label $1\sim 4$ in Figure~\ref{fig:AMR-demo}, and each group consists of 256 threads rendering the $16\times 16$ blocks. The tiles are labeled with accuracy levels $1\sim 4$ according to visual eccentricity from the gaze center. If, for instance, a tile is labeled as level 2, then only 2 out of the 4 groups of 256 threads will run through the 3DGS calculation, while the other 2 will exit immediately after checking, thus saving the computing resources considerably. }


\renewcommand{\algorithmicrequire}{\textbf{Input:}}
\renewcommand{\algorithmicensure}{\textbf{Output:}}

\subsection{Summary }
\label{sec:summary_methods}

The A3FR framework seamlessly integrates the techniques discussed in the previous sections. We employ the A3FR-ViT with an early-exit mechanism for eye-tracking, which generates a sequence of increasingly accurate gaze predictions at multiple exit points. The GPU implements 3DGS to perform foveated rendering based on these gaze predictions in an incremental fashion, with rendered regions dynamically adjusted as predictions improve. The rendering process is adaptive, meaning the resolution of tiles varies based on the number of gaussians intersecting with each tile. Our AMR strategy enables progressive refinement of the output image over multiple rendering rounds. The foveated rendering aspect is achieved by adjusting rendering resolution according to the eccentricity from the user's gaze point. The incremental rendering approach effectively balances the rendering and eye-tracking workloads between CPU and GPU resources. The complete A3FR framework is outlined in Algorithm~\ref{alg:parallel_render}.

\section{Experiments}
\label{sec_exp}

\subsection{Settings}
\label{sec:setting}
\paragraph{Training} We use the original 3D Gaussian Splatting~\cite{kerbl20233d} as our baseline for full-resolution rendering. Following the established methodology, we use L-1 loss and D-SSIM regularization, and incorporate both pruning and densification strategies during the optimization phase as specified in the original work~\cite{kerbl20233d}. For our experiments, all scenes undergo consistent training for 30,000 iterations to ensure fair comparison across different rendering approaches.

The A3FR-ViT architecture consists of $N=6$ transformer layers, with a local exit implemented at the end of each layer. The baseline ViT model maintains the same $N=6$ layer structure with early exits at each layer ($n=1$). For training, we compute loss at each exit point using a batch-based maximum loss approach. More specifically, in Equation \ref{eqn:multiresolution-loss}, for training dataset $D$ partitioned into batches $B$, the loss $L_g$ is calculated as $\sum_{b \in B} \max_{d \in D_b} (|| \theta_d - \theta_{gt} ||^2)$, where $\theta_d$ represents the predicted gaze direction and $\theta_{gt}$ represents the ground truth.

To evaluate A3FR-ViT, we compared it with ResNet-34~\cite{resnet_inception} and DeepVOG~\cite{deepvog}. For ResNet-34, we added early exit points after each of its four residual blocks by connecting intermediate features to linear layers for gaze prediction. With DeepVOG, we maintained its encoder-decoder structure and positioned exit points after the encoding stream and after each upsampling layer in the decoding stream. All models were trained using the same loss function to ensure fair comparison.

\paragraph{Datasets} For 3DGS training, we adopt four scenes from Tanks\& Temples~\cite{tandt} and Deep Blending~\cite{deep_blending} datasets. All the eye-tracking models are trained and evaluated on the OpenEDS datasets~\cite{garbin2019openeds,emery2021openneeds}, which were captured by VR device cameras in real time. OpenEDS data are eye images of 640$\times$400 resolution and are grouped into 9160 sequences of continuous eye movement, with each sequence containing 100 images. All of the gaze tracking networks are trained with 25 epochs.

\begin{algorithm}[t]
\caption{A3FR Framework}
\label{alg:parallel_render}
\begin{algorithmic}[1]
    \Require   $V(.)$ \Comment{A3FR-ViT function, return series $V_l$ at layer $l$}
    \Require  $G(.)$ \Comment{A3FR-3DGS function, return rasterized pixels}
    \Require $E$ \Comment{User's eye image captured}
    \Require  $N$ \Comment{Max layers of ViT}
    \State $r \gets$ Profiling($V$(dataset)) \Comment{ Eq.~\ref{eqn:criteria1}: foveal regions}
    \State $S \gets $ 0, (0,0) \Comment{Init shared variable: layer index, gaze center} 
    \State \textbf{Spawn Process 1 and Process 2 in Parallel}
    \Statex  \textbf{Process 1:} \Comment{Gaze tracking}
    \For{ $l =1$ \textbf{to} $N$}
    \State $(x,y) \gets V_l(E) $ \Comment{Run ViT forward by 1 layer}
    \State $S \gets l, (x,y)$ \Comment{Record gaze tracking result}
    \EndFor
    \Statex \textbf{End Process 1}
    \setcounter{ALG@line}{3}
    \Statex \textbf{Process 2:} \Comment{Rendering}
    \State $I \gets 0$ \Comment{Init Canvas}
    \State $l, (x,y) \gets S$ \Comment{Query gaze tracking result}
    \While{$l\leq N$ }
    \State $M\gets$ AdaptiveMesh($l$, $(x,y)$, $r$) \Comment{AMR}
    \State $I \gets G(M)$ \Comment{3DGS rasterization using mesh M} 
    \State $l, (x,y) \gets S$ \Comment{Update gaze tracking result}
    \EndWhile
    \Statex \textbf{End Process 2}
\end{algorithmic}
\end{algorithm}

\paragraph{Hardware} The experiments were conducted on a machine with 4-core 8-thread Intel Xeon Platinum 8259CL CPU and an Nvidia Tesla T4 16GB GPU (1590 MHz clock, 2560 CUDA cores). The experiments were conducted under two conditions: the default settings and a modified setting with a 915 MHz clock frequency and 1024 CUDA cores. This adjustment was made to mimic the performance of an edge device GPU, specifically the Jetson Orin NX 16GB, which operates at a 918 MHz clock frequency and possesses 1024 CUDA cores~\cite{nvidia-gpu}, and has been frequently used in prior research to model rendering performance in VR devices \cite{gilles2023holographic, zhang2024boxr, zhang2024co, pancrisp, he2024omnih2o, vulkansim}.
Similarly, two settings are applied for CPU clock frequency, 2.5 GHz and 2.2 GHz, where 2.5 GHz is the default CPU frequency, and 2.2 GHz is used to simulate the clock frequency for Jetson Orin NX 16GB. 


 
\paragraph{Baseline comparison}
The default A3FR framework loads the 3DGS rendering on GPU and the eye-tracking network on CPU in parallel with communication. In experiments we compare A3FR against two baseline rendering approaches.
\begin{itemize}
    \item The original 3DGS rendering on GPU without the using the gaze tracking mechanism, denoted as \textit{Full Resolution Rendering (FRR)}.
    \item The traditional TFR framework where the gaze tracking process and the foveated rendering process are executed sequentially, denoted as~\textit{Sequential Foveated 3DGS Rendering (SFR)}.
\end{itemize}

For all the algorithms, the tiling size in the 3DGS process is set at 32$\times$32. Evaluations are conducted at three different resolutions: 1280$\times$720, 1920$\times$1080, and 2560$\times$1440 (720p, 1080p, and 1440p).

\subsection{Performance on Gaze Tracking}
In this section, we assess the performance of gaze tracking by comparing A3FR-ViT with two other gaze tracking neural networks. For A3FR-ViT, we enhance efficiency by applying tokenwise pruning, which eliminates redundant tokens that have low attention scores, thereby reducing the computational cost of gaze tracking. Specifically, we implement two pruning ratios, $10\%$ and $20\%$, to evaluate the impact of this reduction on the performance of A3FR-ViT.


Table~\ref{tab:gaze_err} presents a summary of gaze tracking errors on the test set of the OpenEDS dataset, represented by $\sigma_x$ and $\sigma_y$ in degrees. It also details the processing latency on the Jetson Orin NX 16GB CPU. For the A3FR-ViT model, the embedding layer consists of a single convolutional layer that transforms the input image into a sequence of tokens. In contrast, for the ResNet-based models, the embedding layer comprises a convolutional layer, a batch normalization layer, and a ReLU layer. For A3FR-ViT, ResNet-based models, and DeepVoG, local exits are introduced to generate the gaze location at the end of each block. Specifically, six local exits are used for A3FR-ViT and DeepVoG, while the ResNet-based models utilize four local exits.

From Table~\ref{tab:gaze_err}, several key observations emerge. First, the gaze tracking error decreases for local exits attached to the deeper stages of the neural network, as more layers contribute to processing the gaze data, resulting in more accurate predictions. Second, A3FR-ViT demonstrates significantly lower gaze tracking latency on the VR CPU compared to the other two baselines, while maintaining comparable accuracy. This improvement is attributed to the implementation of tokenwise pruning and the inherent accuracy advantages of Vision Transformers (ViT) over traditional CNNs. Finally, as the pruning ratio increases, the gaze tracking latency for A3FR-ViT significantly decreases without a substantial loss in accuracy. For instance, with $20\%$ of tokens pruned, A3FR-ViT achieves an end-to-end latency of 26.28ms. Increasing the pruning ratio to $20\%$ further reduces the latency to 21.64ms, demonstrating the effectiveness of pruning in enhancing processing efficiency while retaining performance.

\begin{figure}
    \centering
    \includegraphics[height=0.41\linewidth]{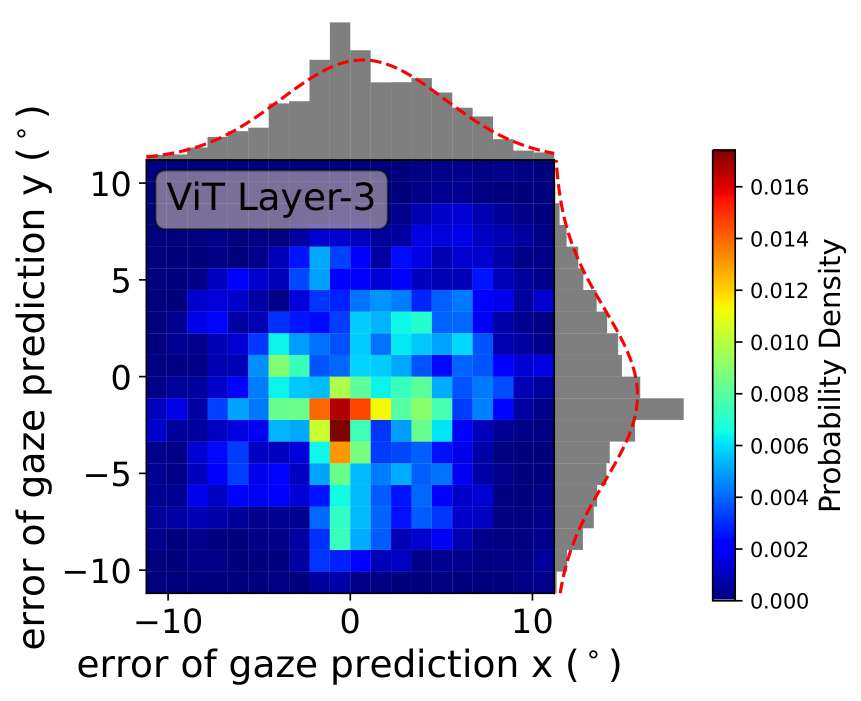}
    \includegraphics[height=0.41\linewidth]{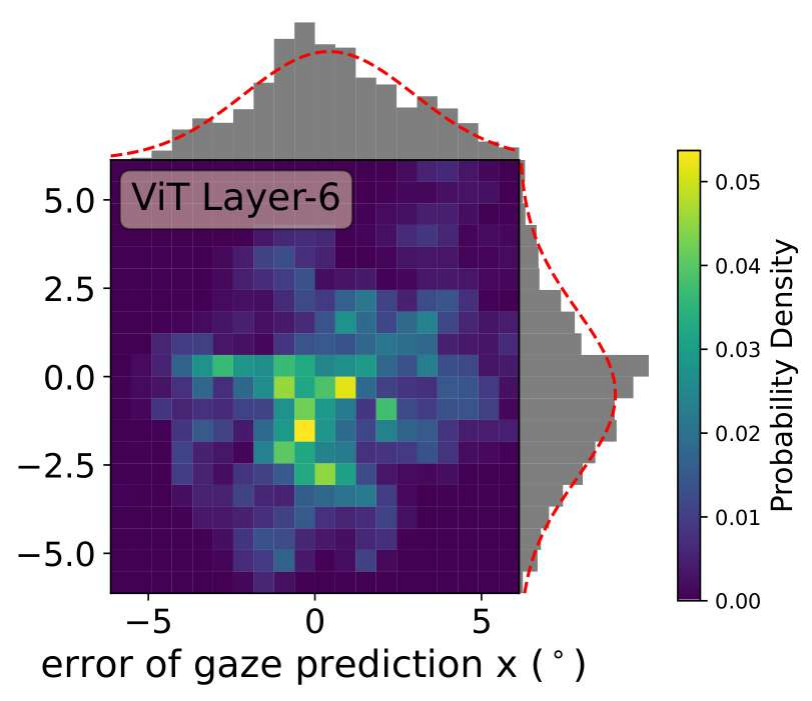}

    \caption{Left: Joint and marginalized distribution of gaze tracking error using the early exit results at 3rd layer of A3FR-ViT. Right: The error distribution of the final gaze prediction at 6th layer. }
    \label{fig:gaze_error}
\end{figure}

During our offline profiling as Described in Sec. \ref{sec:a3fr_scheme}, both the average tracking error and its distribution are needed. We used the error tested on OpenEDS dataset as the underlying probability distribution for profiling. As an example, Figure ~\ref{fig:gaze_error} shows the 2D probability distribution, marginalized distribution and Gaussian fit of gaze tracking error for A3FR-ViT at layer-3 and layer-6. 

The lower two rows for each model in Table~\ref{tab:gaze_err} report the by-layer and cumulative latencies. We find that pruning gives moderate speedup while slightly influencing tracking accuracy. The ResNet-based model and DeepVOG are more expensive computationally, giving significant delays of latency.

\begin{table*}
  \caption{Accuracy Evaluation on Gaze Tracking Performance}
  \label{tab:gaze_err}
  \resizebox{0.85\textwidth}{!}{%
  \begin{tabular}{ccccccccl}
    \toprule
    & & Embedding& Layer-1& Layer-2& Layer-3& Layer-4& Layer-5& Layer-6 \\
    \midrule
    & Error $(\sigma_x $, $\sigma_y)/^\circ$ & $-$ &(8.40, 9.33)  &(6.08, 7.05)  &(4.50, 4.46)  &(3.58, 3.38)  &(2.97, 2.85)  &(2.05, 2.16) \\
    A3FR-ViT & Latency/ms & 1.00 & 4.10 & 4.34 & 4.46 & 4.40 & 3.90 &  4.06  \\
    & Cumulative Latency& 1.00 & 5.11 & 9.45 & 13.91 & 18.31 & 22.21 & 26.28 \\
    \midrule
    & Error $(\sigma_x $, $\sigma_y)/^\circ$ &$-$ &(8.19, 9.22)  &(6.10, 6.98)  &(4.40, 4.53)  &(3.55, 3.76)  &(3.01, 2.96)  &(2.53, 2.56)\\
    A3FR-ViT  & Latency/ms & 0.93 & 3.73 & 4.21 & 3.98 & 3.76 & 3.91 & 3.70\\
    (0.1 pruned)& Cumulative Latency&0.93 & 4.66 & 8.88 & 12.85 & 16.62 & 20.52 & 24.23\\
    \midrule
    & Error $(\sigma_x $, $\sigma_y)/^\circ$ &$-$  &(8.17, 9.62)  &(6.07, 6.85)  &(4.61, 4.42)  &(3.74, 3.55)  &(3.16, 2.88)  &(2.69, 2.39) \\
    A3FR-ViT  & Latency/ms & 1.07 & 3.62 & 4.56 & 3.40 & 3.06 & 2.99 & 2.92\\
    (0.2 pruned)& Cumulative Latency& 1.07 & 4.70 & 9.26 & 12.66 & 15.72 & 18.72 & 21.64\\
    \midrule
    & Error $(\sigma_x $, $\sigma_y)/^\circ$ &$-$ &(7.96, 10.04)  &(5.45, 5.70)  &(2.71, 2.38)  &(2.24, 1.94) & $-$ & $-$\\
    ResNet-based & Latency/ms & 3.71 & 6.48 & 6.86 & 11.63 & 8.61& $-$ & $-$ \\
    & Cumulative Latency& 3.71 & 10.19 & 17.05 & 28.69 & 37.30& $-$ & $-$\\
    \midrule
    & Error $(\sigma_x $, $\sigma_y)/^\circ$  & $-$ &(6.27, 8.24)  &(4.34, 5.13)  &(3.19, 3.53)  &(2.63, 2.88)  &(2.24, 2.42)  &(1.91, 2.03)\\
    DeepVOG  & Latency/ms& $-$ & 10.06 & 6.78 & 24.77 & 28.55 & 35.31 & 9.13   \\
    & Cumulative Latency& $-$& 10.06 & 16.84 & 41.61 & 70.16 & 105.47 & 114.6   \\
  \bottomrule
\end{tabular}
}
\end{table*}

\subsection{A3FR Latency Evaluation}
\label{sec:latency-eval}
In this section, we assess the latency performance of A3FR by comparing it with two baseline methods, FRR and SFR, as described in Section~\ref{sec:setting}. Specifically, we evaluate all three solutions on four scenes—"truck," "train," "drjohnson," and "playroom"—sourced from the Tanks \& Temples~\cite{tandt} and Deep Blending~\cite{deep_blending} datasets. The evaluation is conducted at three different resolutions: $1280\times 720$, $1920\times 1080$, and $2560\times 1440$ for each scene, with 100 camera poses per resolution. The average rendering latency across these 100 camera poses is recorded. The experiments are performed using the hardware setup detailed in Section~\ref{sec:setting}.

\begin{figure}
    \centering
    \includegraphics[width=\linewidth]{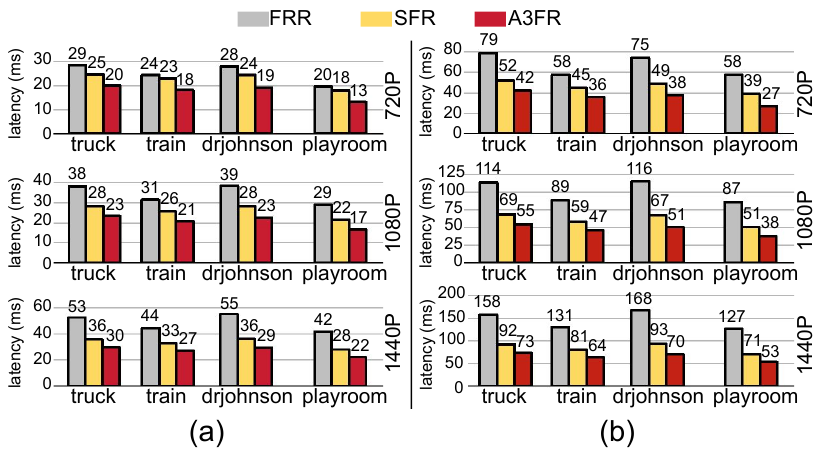}
    \caption{(a) Comparison of rendering latency for A3FR against other baseline models across 4 scenes and 3 resolutions. Gray columns represent FRR with the original 3DGS model. Yellow columns represent SFR where eye-tracking ViT and 3DGS are running on the same GPU in serial. Red columns represent A3FR. Latencies are shown in (ms). (b) Latency evaluation encompasses three approaches, with the GPU clock frequency and CUDA cores adjusted to align with the Jetson NX Orin 16GB.}
    \label{fig:latency_T4}
\end{figure}

The results are presented in Figure~\ref{fig:latency_T4}, where Figure~\ref{fig:latency_T4} (a) illustrates the performance under standard CPU and GPU configurations, while Figure~\ref{fig:latency_T4} (b) shows the results on the Jetson Orin NX. Overall, A3FR consistently outperforms the baseline methods across all tested scenes, resolutions, and hardware setups. At 1080p and 1440p resolutions, A3FR achieves an average speedup of $20\%$ compared to SFR and $40\%$ compared to FRR. For 720p, the performance gain of SFR over FRR is relatively smaller due to eye-tracking overhead, whereas A3FR still delivers an average speedup of $30\%$. The results remain consistent across different scenes. This is because A3FR parallelizes eye tracking with the 3DGS process by distributing them between the CPU and GPU, significantly reducing overall computational latency. Additionally, we evaluated rendering performance on the "playroom" scene at 720P resolution under reduced settings equivalent to 16 TOPS. The results demonstrate a of latency of 151.239 ms, 104.328 ms and 76.757 ms for FRR, SFR and A3FR, respectively. 

To analyze the details of latency improvement, we present a breakdown of latency for a sample input frame (Figure~\ref{fig:breakdown_combined} (a)). Specifically, we select the "truck" scene at 1080p resolution with 100 camera poses. Let \( R_i \) denote the rendering process of $C_{i}$, as illustrated in Figure~\ref{fig:render_rounds}, and let \( L_i \) represent the computational latency on the CPU at the \( i \)-th exit point. It is important to note that the maximum index \( i_{\text{max}} \) for \( L_i \) and \( R_i \) may differ due to variations in processing speed, causing either gaze tracking or 3DGS rendering to complete fewer rounds. We observe that the preprocessing stage of the 3DGS rendering pipeline requires a considerable amount of computation, which can be initiated before the gaze tracking process begins. Within the A3FR framework, the CPU and GPU synchronize to enable parallel processing, effectively reducing overall rendering latency.

\subsection{Impact of Gaze-tracking Network}
\label{sec:ablation:gaze-tracking}
\begin{figure*}[t]
    \centering
    \includegraphics[width=\linewidth]{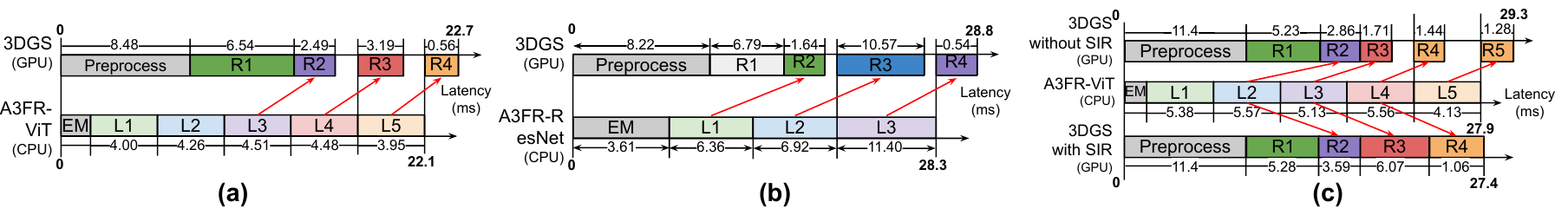}
    \caption{(a) Latency breakdown of A3FR framework. For 3DGS, the gray bar represents the preprocessing latency, while the colored bars labeled R1 to R4 indicate the latencies for each rendering round. Both individual and cumulative latencies are displayed. For ViT, the gray bar represents the patch embedding latency, and the colored bars labeled L1 to L5 correspond to the processing latency at each local exit point. Red arrows illustrate the dependency of each 3DGS rendering round on the results from the corresponding local exit of A3FR-ViT. (b) Latency breakdown with ResNet-based Model as the gaze tracking network. (c)  Impact of speculative incremental rendering on the overall latency.}
    \label{fig:breakdown_combined}
\end{figure*}

In this section, we analyze the impact of A3FR-ViT on overall rendering latency. Specifically, we replace A3FR-ViT with the ResNet-based gaze tracking model from Table~\ref{tab:gaze_err} and repeat the latency measurement on the "truck" scene at 1080p resolution over 100 camera poses. The breakdown is presented in Figure~\ref{fig:breakdown_combined} (b). Compared to A3FR-ViT, the ResNet-based model significantly increases the overall TFR processing time. This is primarily because L3 is considerably more time-consuming ($\sim$11ms) than a ViT layer ($\sim$4ms), while early predictions from L1 and L2 exhibit high error rates, making the intermediate layers of ResNet-based models tend to be either computationally expensive or less accurate.

\subsection{Effect of Speculative Incremental Rendering}
\label{sec:ablation:incremental_rendering}
In this section, we examine the impact of A3FR latency caused by variations in the relative processing speeds of the CPU and GPU. To simulate performance fluctuations, we use~\textit{systemd} to consume CPU computational resources, thereby slowing down the processing speed of A3FR-ViT during the TFR process. The CPU utilization rate is alternated between 50\% and 100\% at 0.1-second intervals. All evaluations are conducted on the "truck" scene at 1080p resolution and repeated over 100 camera poses. We compare the A3FR against a baseline algorithm that did not contain speculative incremental rendering. Namely, if the current rendering round is finished and new gaze tracking results has been received by GPU, it will wait until receive it before rendering further. 

The results presented in Figure~\ref{fig:breakdown_combined} (c) demonstrate that the A3FR framework remains resilient to performance fluctuations. As shown in Figure~\ref{fig:breakdown_combined} (a), a reduced processing speed in A3FR-ViT leads to slower gaze tracking output. However, compared to A3FR without speculative incremental rendering, it still achieves a relatively low overall TFR latency of 26.0ms. In contrast, without speculative incremental rendering, the next rendering round would only begin upon receiving the subsequent gaze tracking result, significantly increasing the overall rendering latency.


\subsection{Impact of AMR}
\label{sec:ablation:amr}
\begin{figure}
    \centering
    \includegraphics[width=\linewidth]{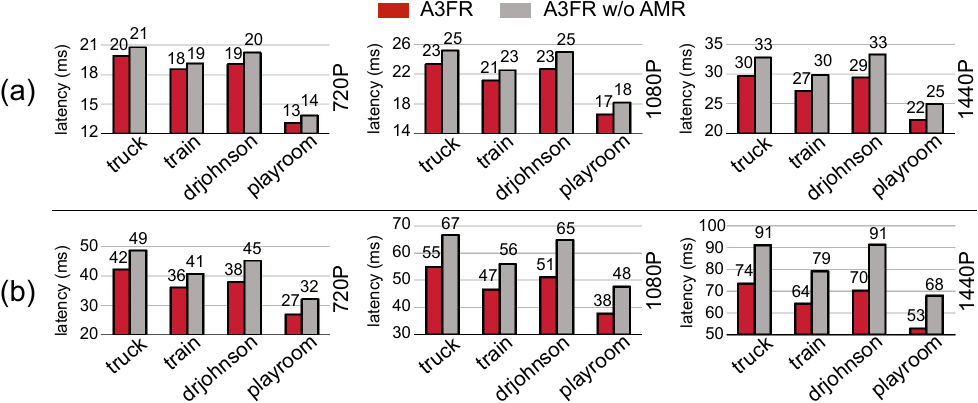}
    \caption{(a) Ablation study of AMR strategies. A3FR rendering latencies are shown in red and those without AMR are shown in grey. (b) Same as (a), but run on the machine with reduced GPU clock frequency and CUDA cores to match Jetson NX Orin 16GB. }
    \label{fig:AMRset_Jetson}
\end{figure}

Finally, we evaluate the impact of AMR on reducing TFR rendering latency. To achieve this, we remove AMR from the A3FR framework and repeat the latency evaluation across four scenes and three different resolutions, using the same experimental settings as described in Section~\ref{sec:latency-eval}. The results are presented in Figure~\ref{fig:AMRset_Jetson}. We observe that AMR contributes to approximately a 10\% improvement in rendering latency at 1440p resolution, while its impact is less pronounced at lower resolutions. This is primarily due to the additional computational overhead introduced by AMR. Specifically, during preprocessing, tiles must be sorted based on the number of Gaussians, and their resolutions adjusted accordingly. When the rendering load is relatively low, such as in the 720p case, this additional processing cost becomes comparable to the potential savings, reducing the overall benefit of AMR.

\subsection{User Study}
\label{sec:user_study}
To evaluate the practical effectiveness of our A3FR approach, we conducted a comprehensive user study to assess the visibility of artifacts in foveated 3D Gaussian Splatting rendering. The primary objective of the study was to quantitatively assess the rendering quality achieved by our A3FR framework (Sections \ref{sec:a3fr_scheme} and \ref{sec:AMR}) and to validate its potential to deliver high-quality visual experiences under gaze-tracked conditions. This objective was met by demonstrating that participants could not reliably distinguish between the A3FR and full-resolution rendering methods, confirming that the overall perceptual quality remains uncompromised.

\begin{figure}[t]
\centering
\begin{minipage}{0.45\linewidth}
    \centering
    \includegraphics[width=0.95\linewidth]{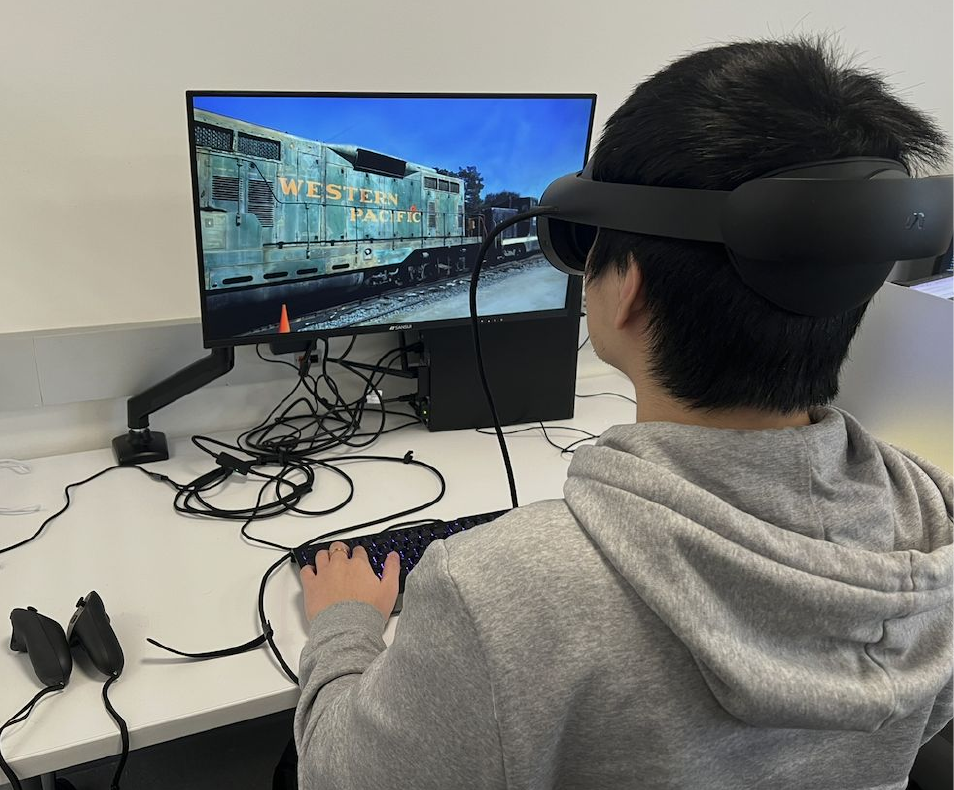}
    \caption{Participants are conducting the user study on the Quest Pro.}
    \label{fig:userstudy-photo}
\end{minipage}%
\hfill
\begin{minipage}{0.5\linewidth}
    \centering
    \includegraphics[width=\linewidth]{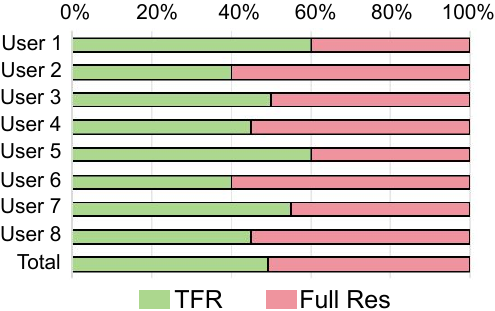}
    \caption{Selection results from the participants.}
    \label{fig:userstudy-results}
\end{minipage}
\end{figure}

A total of eight participants were recruited for the experiment, ensuring a representative sample for evaluating the method. As shown in Figure \ref{fig:userstudy-photo}, during the experiment, each participants remained seated and observed the stimuli via a HMD, the Meta Quest Pro~\cite{meta_quest}. Interaction during the study was facilitated through a standard keyboard interface, providing a uniform and straightforward means of navigating between stimuli.
The stimuli comprised images rendered from various scenes using 3DGS. Participants were instructed to perform a two-interval forced-choice (2IFC) task~\cite{yeshurun2008bias}, a widely used methodology for assessing visual preferences. In each trial, participants were presented with rendered images applied to the same scene: (1) the reference image, from the ground truth dataset of 3DGS, and (2) two test images, labeled t1 and t2, rendered using different methods.
One of the test images is generated using our A3FR approach with AMR, which incorporates a pre-defined fixed point to simulate the gaze direction. The other test image depicts a fully rendered scene using 3DGS without gaze-tracking. These two rendering methods, referred to as m1 (A3FR) and m2 (FRR), provides a basis for evaluating the impact of TFR on perceived visual quality.

During each trial, t1 and t2, which depicted the same visual scene, were randomly paired with m1 and m2 to eliminate order effects or potential bias. Participants were allowed to freely switch among t1, t2, and the reference image using the keyboard, enabling them to carefully assess and compare the visual quality of the test images.

\begin{figure}
    \centering
    \includegraphics[width=0.8\linewidth]{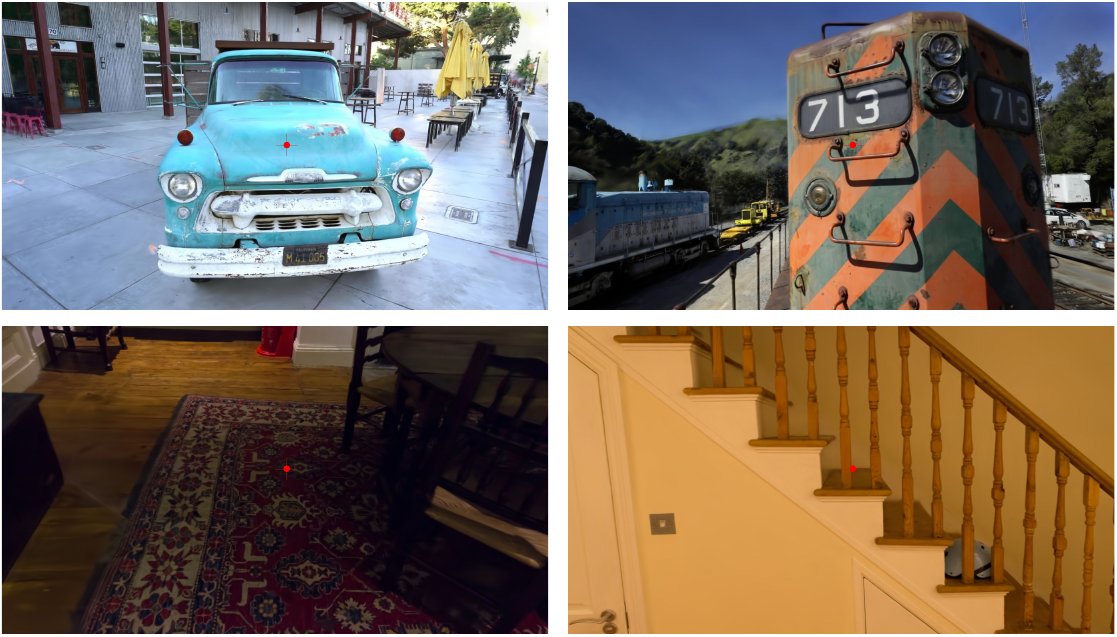}
    \caption{Sample images from 20 views of 4 scenes.}
    \label{fig:userstudy-sample}
\end{figure}

To simulate realistic gaze-tracking conditions, participants were instructed to focus on the marked pre-defined fixed point (as shown in Figure \ref{fig:userstudy-sample}) throughout the comparison and were required to make their judgments within a 10-second time frame. As noted in Section ~\ref{sec:bg:human-eye}, the typical duration of an eye fixation ranges from tens of milliseconds to several seconds \cite{fixduration}. To accommodate the brief delay caused by switching between image pairs and the subsequent visual recovery period, we extended the fixation duration to 10 seconds as the trial time. These constraints were designed to ensure that the decision-making process closely mirrored the time-sensitive dynamics of real-time gaze-tracked rendering systems.

After observing both test images at least once, participants were asked to select the test image they perceived as having higher visual quality. A total of 20 distinct image pairs (sample images shown in Figure~\ref{fig:userstudy-sample}), rendered from 4 different scenes, were presented to each participant, resulting in 20 trials per participant. This design provided a robust dataset for analyzing user preferences and visual performance across various scenarios.
The results of the user study are illustrated in Figure~\ref{fig:userstudy-results}. Across all participants, the TFR within our A3FR framework was selected 49.4\%$\pm$8.2\% of the time compared to the baseline rendering method, demonstrating that our rendering method maintains visual quality and user experience without significant degradation compared to the baseline method, effectively avoiding perceivable decline in image quality or overall visual satisfaction. This outcome underscores the practicality of the A3FR framework for real-world applications in foveated rendering systems.

\section{Conclusions}
\label{sec:conclusion}
In this work, we propose a collaborative execution framework, A3FR, that performs foveated rendering and gaze tracking in parallel to reduce the overall processing latency in AR/VR systems. We further introduce A3FR-ViT, an efficient gaze-tracking neural network that enables early estimation of gaze direction and facilitates parallel processing with 3DGS rendering.
A3FR significantly reduces rendering latency while maintaining equal visual quality validated by our user study. Experimental results demonstrate its performance across various software and hardware settings. This framework paves the way for more efficient and responsive rendering systems in VR applications.

\newpage
\bibliographystyle{ACM-Reference-Format}
\bibliography{refs}

\end{document}